\documentstyle[twocolumn,prl,aps,epsf]{revtex}
\large
\begin{document}
\epsfverbosetrue
\draft
\input epsf.sty
\twocolumn[
\hsize\textwidth\columnwidth\hsize\csname @twocolumnfalse\endcsname
\title{X-ray Dichroism and Orbital Anapoles}
\author{Paolo Carra$^1$, Andr\'{e}s Jerez$^2$\cite{AAuth}, 
and Ivan Marri$^1$\cite{AAAuth}} 
\address{$^1$European Synchrotron Radiation Facility,
B.P. 220, F-38043 Grenoble C\'{e}dex, France\\
$^2$Institut Laue-Langevin, B.P. 156X, F-38042 
Grenoble C\'{e}dex, France}
\date{\today} 
\maketitle 
\begin{abstract}
We present a theoretical analysis of photo-absorption spectroscopies in
noncentrosymmetric systems, covering both x-ray and optical regions.
Integrated dichroic spectra are interpreted using microscopic effective operators, 
which are obtained by coupling the orbital angular momentum to the 
orbital anapole. Symmetry arguments afford a 
classification of valence-electron states in the presence of parity
nonconserving hybridisation. Enantiomorphism is identified by a 
two-particle chiral operator.
\end{abstract}
]
The current paper aims at providing a theoretical interpretation of 
two recent experiments: 
\begin{itemize}
\item[] {\it X-ray natural circular dichroism} (XNCD),
probed in $\alpha$-LiIO$_3$ \cite{Gou98} and in 
Na$_3$Nd(digly)$_3\cdot$2NaBF$_4\cdot$6H$_2$O \cite{Ala98}. 
(The effect was observed at the iodine $L$ edges 
and at the Nd $L_3$ edge.)
\item[]  {\it X-ray nonreciprocal linear dichroism} (XNRLD),
detected at vanadium $K$ edge in the low-temperature 
insulating phase of a Cr-doped V$_{2}$O$_{3}$ crystal\cite{Gou00}.
\end{itemize}

XNCD measures the difference in absorption between right 
and left circularly polarised radiations. 
XNRLD implies a difference in absorption between 
radiations with linear polarisation parallel or perpendicular to a local
symmetry axis. Both phenomena stem from the interference 
between electric-dipole (E1) and electric-quadrupole (E2) transitions 
that raise an inner-shell electron to empty valence orbitals. 
Detecting a nonvanishing signal thus requires
an ordered structure (crystal) and the breaking of space inversion. 

As will be shown, Lie groups (or, alternatively, Lie algebras) furnish 
a powerful tool for 
interpreting x-ray dichroism in non-centrosymmetric crystals. 
In fact, effective microscopic operators, which express electronic
properties revealed by the spectra, are readily deduced 
from the pertinent group generators: {\em The orbital angular momentum 
and orbital anapole}. 
Our approach hinges therefore on spectrum-generating algebras (SGAs), 
a concept originally introduced in nuclear \cite{GoL59} 
and particle physics \cite{DGN65,DaN66}. 

We can switch from such an abstract formulation to a 
physical picture that is easy to grasp. This is achieved 
by relating the chosen SGA to the electron 
orbital current. To this end, Trammell's 
expansion\cite{Tra53} is exploited. 

As no pseudoscalar is accessible to the E1-E2 interference,
we must include E1-magnetic-dipole (M1) excitations in order 
to discuss orbital chirality.
Our approach readily extends to the optical range.   
Results for the E1-M1 circular dichroism (observable also in powder samples) 
will thus be reported.

Our analysis is restricted to dichroic spectra integrated 
over a finite energy range, corresponding to the two partners
of a spin-orbit split inner-shell: $j_{\pm}=l_c\pm \frac{1}{2}$.

We follow Carra and Benoist\cite{CaB00} and consider the 
general framework of a de Sitter algebra: 
$so(3,2)$ \cite{GoL59,Eng72}.
A realisation of such an algebra is provided by the 
operators:  
${\bbox A}^-=i({\bbox A}-{\bbox A}^{\dagger})/2$, 
${\bbox A}^+=({\bbox A}+{\bbox A}^{\dagger})/2$, 
${\bbox L}$ and $N_0$, 
where
$
{\bbox A}={\bbox n}\,f_1(N_0)+{\bbox \nabla}_{\Omega}f_2(N_0)\, ,
%
$
with ${\bbox n}={\bf r}/r$ and 
${\bbox \nabla}_{\Omega}=-i{\bbox n}\times{\bbox L}$; 
${\bbox L}$ denotes the orbital angular momentum (in units 
of $\hbar$). Furthermore, 
$
f_1(N_0)=(N_0 - 1/2)f_2(N_0)
$
and 
$
f_2(N_0)=\sqrt{(N_0-1)/N_0},
$
where 
$N_0|lm\rangle=(l+\frac{1}{2})|lm\rangle$, with $|lm\rangle$ a 
spherical harmonic. ${\bbox A}$ and ${\bbox A}^{\dagger}$ 
are known as shift operators as their action on $|lm\rangle$ 
changes $l$ into $l\pm1$.

As will be seen, discussing x-ray dichroism in noncentrosymmetric systems 
only requires ${\bbox L}$ (rotations) and ${\bbox A}^-$ (boosts), 
which themselves generate a homogeneous Lorentz group: $SO(3,1)$. 
We observe that 
\begin{equation}
{\bbox L} \cdot {\bbox A}^- = {\bbox A}^- \cdot {\bbox L} = 0\, ,
\label{ortho} 
\end{equation}
and that  
a physical interpretation of ${\bbox A}^-$
is provided by the relation 
\begin{eqnarray}
{\bbox \Omega}^- &=&
({\bbox n}\times{\bbox L}-{\bbox L}\times{\bbox n})/2
=i\left[{\bbox n},{\bbox L}^2\right]/2
\label{anapole}
\\
&=&\frac{i}{2}\left({\bbox \nabla}_{\Omega}^{\dagger} - 
{\bbox \nabla}_{\Omega}
\right)
=\frac{1}{2}\frac{1}{\sqrt{N_0}}\left[ N_0, {\bbox A}^- 
\right]_+\frac{1}{\sqrt{N_0}}\, ,
\nonumber
\end{eqnarray}
where $[...]_+$ denotes an anticommutator. 
Eq. (\ref{anapole}) defines the (purely angular) 
orbital anapole \cite{Zel57,FaK80,BaB97}: ${\bbox \Omega}^-$.
As pointed out by Dothan {\it et al.}\cite{DaN66}, 
${\bbox L}$ and ${\bbox \Omega}^-$ also generate $SO(3,1)$. 
[${\bbox L}$ and ${\bbox A}^+$ generate another $SO(3,1)$ 
subgroup of $SO(3,2)$. Being space-odd and time-even, ${\bbox A}^+$ 
does not enter our theory of dichroism in noncentrosymmetric 
systems.]

Our formulation is based on a localised (atomic) model 
and exploits the Racah-Wigner calculus to relate 
integrated dichroic spectra to the ground-state expectation value of 
effective orbital operators.  These operators are constructed by 
coupling  the hermitean vector operators ${\bbox L}$ (space-even) 
and ${\bbox \Omega}^-$ (space-odd) 
to obtain space-odd irreducible tensors. In such a framework, 
one-electron tensors are given by 
($\iota$ is an electron label)
\begin{eqnarray}
&&
\sum_{\iota} [{\bbox L}_{\iota}\otimes
{\bbox \Omega}_{\iota}^-]^{(k)l\leftrightarrow l'}_q
\equiv
\sum_{\iota} [{\bbox L}_{\iota}\otimes
{\bbox \Omega}_{\iota}^-]^{(k)l,l'}_q
+{\rm H.c.}
\nonumber
\\
&&
=\sum_{mm'}\langle l'm' |
[{\bbox L}\otimes{\bbox \Omega}^-]^{(k)}_q|lm \rangle 
a^{\dagger}_{l'm'}a_{lm}+{\rm H.c.}
\label{one_part}
\end{eqnarray}
Here, $l'=l\pm1$; $a^{\dagger}_{l m}$ and $a_{l' m'}$ create and annihilate 
valence electrons. (As usual, tensor couplings are
defined via Clebsch-Gordan coefficients: 
$
[U^{(p)}\otimes V^{(\kappa)}]^{(k)}_q\equiv
\sum_{\mu,\nu}C^{kq}_{p\mu;\,\kappa\nu}U^{(p)}_{\mu}V^{\kappa}_{\nu}.)
$
Extending Eq. (\ref{one_part}) to define two-electron space-odd irreducible
tensors is straightforward. (Point-group symmetry can be added to 
our analysis by introducing 
{\it point-group coordinates} \cite{CaT94} and retaing the pertinent 
totally-symmetric representations. 
Full details will be given elsewhere.)

{\em E1-E2 interference: XNCD}\cite{Gou98,Ala98} The effect is characterised 
by an {\it orbital pseudodeviator}, i.e. a space-odd and 
time-even rank-two irreducible tensor \cite{Nat98}. 
In our formalism, its microscopic expression is given by a 
coupling of ${\bbox L}$ to the orbital anapole; it reads:
$
[{\bbox L}\otimes{\bbox \Omega}^-]^{(2)l\leftrightarrow l'}_q
$.
For the integrated XNCD spectrum, 
we thus find
\begin{eqnarray}
&&
\int_{j_-+j_+}
\frac{\sigma_{\rm XNCD}(\omega)}{(\hbar\omega)^2}
d(\hbar\omega)
=\frac{8\pi^2\alpha}{3\hbar c}(2l_c+1)
\nonumber
\\
&&
\sum_{l=l_c\pm1 \atop l'=l\pm1}
R_{l_cl}^{(1)}R_{l_cl'}^{(2)}\,a_{l'}(l_c,l)\,\,\sum_q \sqrt{\frac{3}{2}}
T^{(2)*}_q({\bbox \epsilon}^-,{\bbox \epsilon}^+,\hat{\bf k})_2
\label{XNCD}
\\
&&
\langle \psi_0 |\sum_{\iota} 
\left[{\bbox L}_{\iota}\otimes({\bbox L}\times{\bbox n}-
{\bbox n}\times{\bbox L})_{\iota} \right]^{(2)l\leftrightarrow l'}_q
| \psi_0 \rangle \,  .
\nonumber
\end{eqnarray}
In the foregoing equation, 
$\sigma_{\rm XNCD}(\omega) = \sigma^{\bbox \epsilon^+}_{\rm X}(\omega) - 
\sigma^{{\bbox \epsilon}^-}_{\rm X}(\omega)$ 
denotes the cross section for E1-E2 circular dichroism in the 
x-ray region ($X$), with
\begin{eqnarray}
&&
\sigma^{\bbox \epsilon}_{\rm X}(\omega) = 2\pi^2\alpha\hbar\omega
[
i\sum_f 
\sum_{\iota\iota'}
\langle g\mid {\bbox \epsilon}^{\ast}\cdot {\bf r}_{\iota}
\mid f\rangle
\label{cross_section}
\\
&&
\langle f\mid {\bbox \epsilon} \cdot {\bf r}_{\iota'}
{\bf k}\cdot {\bf r}_{\iota'} \mid g\rangle 
+ {\rm c.c.}
]
\delta(E_f-E_g-\hbar\omega)\, ;
\nonumber
\end{eqnarray}
$\hbar\omega$, {\bf k} and ${\bbox \epsilon}^{\pm}=\mp(i/\sqrt{2})
({\bbox \epsilon}_1\pm i{\bbox \epsilon}_2)$ represent energy, 
wave vector, and circular polarisations of the photon, 
respectively; $\alpha=e^2/\hbar c$. The radial integrals are given by
$
R^{(L)}_{l_cl}=\int_0^{\infty}\!\!\!\! dr \varphi_{l_c}(r) r^{L+2} 
\varphi_l(r)\, ,
$
where $\varphi_{l_c}(r)$ and $\varphi_l(r)$ denote inner-shell 
and valence radial wave functions, respectively \cite{Note1,Note2}. 

A general expression can be given for the wave-vector and polarisation 
responses used throughout this work. It reads: 
$i^zT^{(k)}_q({\bbox \epsilon},{\bbox \epsilon}^*,\hat{\bf k})_z$, 
where
\begin{eqnarray}
&&
T^{(k)}_q({\bbox \epsilon},{\bbox \epsilon}^*,\hat{\bf k})_z
=\sum_y\sqrt{(2z+1)(2y+1)}(-1)^{1+k}
\\
&&
\left\{
\begin{array}{ccc}
1 & k & y \\
1 & 1 & z\\
\end{array}
\right\}
\sum_{\eta\beta}C^{kq}_{y\eta;1\beta}
\sum_{\alpha\delta}C^{y\eta}_{1\delta;1\alpha}
Y_{1\delta}({\bbox \epsilon}^*)Y_{1\alpha}({\bbox \epsilon})
Y_{1\beta}(\hat{\bf k})\,,
\nonumber
\end{eqnarray}
with $z=1,2$ denoting E1-M1 (imaginary) 
and E1-E2 (real) responses, respectively; $\hat{\bf k}={\bf k}/k$. 
Explicit forms are given in Table \ref{TableI}. 

{\em E1-E2 Interference: XNRLD} \cite{Gou00}. One effective operators 
is found in this case: a space and time odd tensor of rank three, i.e. 
an {\it orbital septor}. This unusual magnetic moment stems from a coupling 
of the orbital anapole with the quadrupolar moment of ${\bbox L}$ and 
takes the form
$
[[{\bbox L}\otimes{\bbox L}]^{(2)}\otimes 
{\bbox \Omega}^-]^{(3)l\leftrightarrow l'}_q\, .
$

The integrated XNRLD spectrum can thus be written as
\begin{eqnarray}
&&
\int_{j_-+j_+}\frac{\sigma_{\rm XNRLD}(\omega)}
{(\hbar\omega)^2}
d(\hbar\omega)=\frac{8\pi^2\alpha}{\hbar c}
(2l_c+1)\sum_{l=l_c\pm1 \atop l'=l\pm1}
R_{l_cl}^{(1)}
\nonumber
\\
&&
R_{l_cl'}^{(2)}
b_{l'}(l_c,l)\,\,\sum_q\left( 
T^{(3)*}_q({\bbox \epsilon}^{\parallel},\hat{\bf k})_2
-T^{(3)*}_q({\bbox \epsilon}^{\perp},\hat{\bf k})_2\right)
\label{XNRLD}
\\
&&
\langle \psi_0 |\sum_{\iota} \left[
[{\bbox L}_{\iota}\otimes{\bbox L}_{\iota}]^2\otimes
({\bbox L}\times{\bbox n}
-{\bbox n}\times{\bbox L})_{\iota}
\right]^{(3)\,l\leftrightarrow l'}_q
 | \psi_0\rangle 
\, ,
\nonumber
\end{eqnarray}
where
$\sigma_{\rm XNRLD}(\omega) = 
\sigma^{\parallel}_{\rm X}(\omega) - 
\sigma^{\perp}_{\rm X}(\omega)$ 
stands for the cross section for E1-E2 linear dichroism; here, 
$\parallel$ and $\perp$ denote two orthogonal linear-polarisation 
states \cite{Note2}. 

Notice that the E1-E2 interference contains a further 
magnetic term, which vanishes in the 
geometry of Eq. (\ref{XNRLD}). It is an {\it orbital vector}, with  
polarisation response $T^{(1)}_q({\bbox \epsilon},\hat{\bf k})_2$.
In our formulation, it is given by 
$\sum_{\iota} \left[({\bbox L}_{\iota}
\times{\bbox n}_{\iota}-{\bbox n}_{\iota}
\times{\bbox L}_{\iota})_q\right]^{l\leftrightarrow l'}$, i.e.
by the orbital anapole. 
This toroidal contribution \cite{Zel57} to the orbital current 
(see below) could be detected by the following
experiment. Absorb linearly polarised x rays at the 'magic angle'   
$\widehat{{\bf k}{\bbox B}}\simeq 39.23^{\circ}$ for parallel and antiparallel 
magnetoelectric annealing \cite{Gou00} of the sample; 
${\bbox B}\equiv$ external magnetic field. Subtract the two spectra. 

Stemming from parity nonconserving 
hybridisation of the valence electrons, the effect should be easier 
to detect than its nuclear counterpart \cite{Woo97}, as the work 
of Ref. \cite{Gou00} appears to indicate.

{\em E1-M1 interference.}
As previously stated, our approach readily extends to cover 
the visible range, where the E1-M1 interference is predominant. 
In this case, the cross section for optical natural circular 
dichroism (ONCD) takes the form: $\sigma_{\rm ONCD}(\omega)
=\sigma^{\bbox \epsilon^+}_{\rm O}(\omega)-
\sigma^{\bbox \epsilon^-}_{\rm O}(\omega)$, with \cite{Note3} 
\begin{eqnarray}
&&
\sigma^{\bbox \epsilon}_{\rm O}(\omega) = \frac{2\pi^2\alpha\hbar}{m}
[
\sum_f 
\sum_{\iota \iota'}
\langle g\mid {\bbox \epsilon}^{\ast}\cdot {\bf r}_{\iota}
\mid f\rangle
\label{cross_section2}
\\
&&
\langle f\mid {\bbox \epsilon} \times 
{\bf k}\cdot {\bbox l}_{\iota'} \mid g\rangle 
+ {\rm c.c.}
]
\delta(E_f-E_g-\hbar\omega)\, ,
\nonumber
\end{eqnarray}
with ${\bbox l}=\hbar{\bbox L}$.

For the integrated ONCD spectrum we obtain
\begin{eqnarray}
&&
\int_{j_-+j_+}
\frac{\sigma_{\rm ONCD}(\omega)}{\hbar\omega}
d(\hbar\omega)
=\frac{4\pi^2\alpha}{mc}\sum_{l,l'=l\pm1}
R^{(1)}_{ll'}R^{(0)}_{l'l'}
\nonumber
\\
&& 
d(l,l')\langle \psi_0|
\left\{\sqrt{\frac{2}{3}}\, 
T^{(0)}_0({\bbox \epsilon}^+,
{\bbox \epsilon}^-, \hat{\bf k})_1
\right.
\label{OCND}
\\
&&
\sum_{\iota\neq \iota'}\left[{\bbox L}_{\iota}
\cdot({\bbox L}\times{\bbox n}
-{\bbox n}\times{\bbox L})_{\iota'}
\right]^{l\leftrightarrow l'}
\!\!
-\sum_q \sqrt{2}T^{(2)}_q({\bbox \epsilon}^+,
{\bbox \epsilon}^-, \hat{\bf k})_1
\nonumber
\\
&&
\sum_{\iota,\iota'}\left[{\bbox L}_{\iota}\otimes({\bbox L}\times{\bbox n}
-{\bbox n}\times{\bbox L})_{\iota'}
\right]^{(2)l\leftrightarrow l'}_q
\left. \vphantom{\sqrt{\frac{2}{3}}}\right\}|\psi_0\rangle\, ,
\nonumber
\end{eqnarray}
with $d(l,l+1)=1/(l+1)$ and $d(l,l-1)=-1/l$. Two irreducible tensors 
are thus associated to ONCD: a two-particle 
{\it orbital pseudoscalar}, which identifies chirality (see below),
and a orbital pseudodeviator; the latter generalises the results 
of Eq. (\ref{XNCD}) to the two-particle case and will not be dissussed 
any further. 
(There is an important difference 
between E1-E2 and E1-M1 integrated dichroic spectra. E1-E2 implies
excitations from inner shells, which are filled in the
ground state and can therefore be 'integrated out'. As a
result, one-electron properties of the valence states are probed 
in the x-ray region. This is not the case in the optical range; 
E1-M1 involves intra-shell valence excitations and two-electron
quantities are measured.) 

{\em Interpretation of the results.} The choice of ${\bbox L}$ 
and ${\bbox \Omega}^-$ as the building blocks of our analysis 
is physically motivated as follows. As both generators are magnetic, 
they are expected to contribute to the electron orbital current, 
a conjecture which is readily verified. 
To this purpose, consider Trammell's expansion 
of the Fourier transform 
of the atomic orbital magnetisation density\cite{Tra53,BaG88} 
\begin{equation}
{\cal L}({\bf k})
=\frac{1}{2}\langle \psi_0 | \sum_{\iota} 
\left[ {\bbox L}_{\iota}\,f({\bf k}\cdot{\bf r}_{\iota}) +
f({\bf k}\cdot{\bf r}_{\iota})\,{\bbox L}_{\iota}\right]| 
\psi_0 \rangle\, ,
\label{Tra}
\end{equation} 
and perform a recoupling with use of \cite{Lov69}:
\begin{equation}
f({\bf k}\cdot{\bf r})=4\pi \sum_{lm}i^l g_l(kr)Y_{lm}({\bbox n})
Y^*_{lm}(\hat{\bf k})\,,
\label{Lov}
\end{equation}
with
$g_l(x)=(2/x^2)\int_0^x\, d\xi \xi j_l(\xi)$,
where $j_l(\xi)$ denotes a spherical Bessel function. 

We have 
\begin{eqnarray}
&&
{\cal L}({\bf k})=  \sum_{\iota} \langle \psi_0 |\, 
\{g_0(kr_{\iota})\,{\bbox L}_{\iota}
+i\,3g_1(kr_{\iota})/2
\nonumber
\\
&&
(
\hat{\bf k}\times
\left({\bbox L}_{\iota}\times{\bbox n}_{\iota}
-{\bbox n}_{\iota}\times{\bbox L}_{\iota}\right)/2
+\cdot\cdot\cdot\,\,\,\}
| \psi_0 \rangle\, ,
\end{eqnarray}
displaying the orbital angular momentum and anapole contributions, 
and thus providing a physical basis to our chosen $so(3,1)$ SGA, 
which is realised by ${\bbox L}$ and ${\bbox \Omega}^-$.

The representations of the homogeneous Lorentz group, which are 
identified by a pair of indexes, $\nu$ and $\rho$, afford a 
classification of electronic states in noncentrosymmetric
systems. Unitary (unirreps) and nonunitary (nonunirreps) irreducible
representations will be considered. 

It is readily seen \cite{Boh93} that ${\bbox \Omega}^-$ 
corresponds to having $\nu=\rho=0$, yielding the unirrep 
$D(\nu=0,\rho=0)=\sum_{l=0,1,...}^{\infty}
\oplus \,\,D^l$ ("supplementary" series), with basis 
$\sum_{l=0,1,...}^{\infty} \oplus\,\, |lm\rangle$, thus confirming that
we are dealing with a spinless case. (Here, $D^l$ 
identifies the representations of the rotation group.) 
Notice that the spherical harmonics are eigenstates not
only of the $so(3)$ invariant (Casimir) ${\bbox L}^2$, but 
also of the corresponding $so(3,1)$ invariants ${\bbox L}^2
-({\bbox \Omega}^-)^2$ and ${\bbox L}\cdot{\bbox \Omega}^-$. The latter,
with eigenvalues $i\nu\rho$,
identifies {\it the chiral operator} and evinces the
deep interweaving of the Lorentz group with chirality. 
As ${\bbox L}\cdot{\bbox \Omega}^-=0$ [Eq. (\ref{ortho})], 
orbital chirality 
cannot manifest itself at the one-particle level. 

We can build a two-particle basis using the same $so(3,1)$ SGA. 
The generators are given by ${\bbox L}_T =
{\bbox L}_1 +{\bbox L}_2$ and ${\bbox \Omega}^-_T =
{\bbox \Omega}^-_1 +{\bbox \Omega}^-_2$, in this case. 
In such a basis, the two-particle states are eigenstates of the $so(3,1)$ 
invariants; in particular of ${\bbox L}_T\cdot{\bbox \Omega}^-_T$, the chiral
operator that emerges from the E1-M1 processes [Eq. (\ref{OCND})]. 
The Clebsch-Gordan series for the direct product of
two D(0,0) representation takes the form \cite{Nai64}:
$
D(0,0)\otimes D(0,0)=\sum_{\nu=0}^{\infty}\int_{-\infty}^{\infty}
D(\nu,\rho) d\mu(\nu,\rho).
$
The direct product thus decomposes into a direct sum of 
representations some of which characterised by $\nu\rho\neq 0$, 
i.e. orbital chirality appears 
naturally when considering two-particle properties. 

From the relations between the $so(3,1)$ Clebsch-Gordan 
coefficients \cite{ARRW70}, it is readily seen that states with 
$\nu\rho\neq 0$ have mixed parity. For a given value 
of $\nu|\rho|\neq 0$, there are two inequivalent representations with 
basis states of the form $|{\rm even}\rangle + i|{\rm odd}\rangle$ 
and $|{\rm even}\rangle - i|{\rm odd}\rangle$, i.e.
with opposite chiralities. Furthermore, for a given pair $(l_1,l_2)$, 
only chiral stes are allowed to have components with angular momenta  
$(l_1\pm1,l_2)$ and$(l_1,l_2\pm1)$ (two-particle hybridisation). 

We can switch to nonunirreps by introducing 
the transformation:
$
{\bbox N}=({\bbox L}+i{\bbox \Omega}^-)/2,
\quad {\rm and} \quad
{\bbox N}^{\dagger}=({\bbox L}-i{\bbox \Omega}^-)/2\,
$
\cite{FaI69,FaG70}.
${\bbox N}$ is a normal operator; $N_q$ and $N_q^{\dagger}$ 
independently obey the Lie algebra of $SU(2)$. Therefore 
$SO(3,1)\cong SU(2)\times SU(2)$. The invariants are given by
${\bbox N}^2$ and $({\bbox N}^{\dagger})^2$, with eigenvalues $j(j+1)$ 
and $j'(j'+1)$, respectively; $j,j'$ take integer or half-integers values. 
These nonunirreps are denoted by $D^{(j,j')}$. The representation space 
is spanned by the $(2j+1)(2j'+1)$ basis vectors $|jm\rangle|j'm'\rangle$ 
(canonical basis). [All nonunirreps for which $j+j'$ is an integer are
true representations of $SO(3,1)$, while those for which $j+j'$ is half-integer 
are double-valued.] Observe that ${\bbox N}$ and ${\bbox N}^{\dagger}$ 
are not independent as they can be interchanged by space inversion. 
This is equivalent to swapping $j\leftrightarrow j'$. Notice that 
${\bbox L}\cdot{\bbox \Omega}^-=0$ implies $j=j'$. Again, orbital chirality 
does not appear at the one-particle level.

To illustrate the use of nonunirreps in constructing 
two-electron states, we resort to the simple case 
$D^{(\frac{1}{2},\frac{1}{2})}
\otimes D^{(\frac{1}{2},\frac{1}{2})}$.
We have
\begin{displaymath}
\Psi^{LL'}_{MM'}(1,2)=
\sum_{{\rm all} \sigma}
C_{\frac{1}{2}\sigma_1;\frac{1}{2}\sigma_2}^{LM}
C_{\frac{1}{2}\sigma'_1;\frac{1}{2}\sigma'_2}^{LM}
\psi^{\frac{1}{2}}_{\sigma_1,\sigma'_1}(1)
\psi^{\frac{1}{2}}_{\sigma_2,\sigma'_2}(2)\,,
\nonumber
\end{displaymath}
with $\psi^{\frac{1}{2}}_{\sigma,\sigma'}\equiv |\frac{1}{2}\sigma\rangle
|\frac{1}{2}\sigma'\rangle$. In this basis, the two-particle 
chiral operator takes the form: 
$
{\bbox L}_T\cdot{\bbox \Omega}^-_T
=2i\left[{\bbox N}(1)\cdot{\bbox N}(2)-
{\bbox N}^{\dagger}(1)\cdot{\bbox N}^{\dagger}(2)
\right]
$.
Given $\Psi^{LL'}_{MM'}(1,2)$, we can set: 
$(sp)P=\Psi^{01}$, $(ps)P=\Psi^{10}$ 
and similarly for $(s^2)S$, $(p^2)S$, $(p^2)P$, $(p^2)D$. It
is straightforward to verify that $(sp)P$ and $(ps)P$ are 
eigenstates of ${\bbox L}_T\cdot{\bbox \Omega}^-_T$ with opposite 
eigenvalues, and that the remaining states are eigenstates of the two-particle 
chiral operator with zero eigenvalue.
A link between orbital chirality and hybridisation of the valence electrons 
is thus established. 

The linear combination
$\Psi_R=\Psi^{01}\oplus \Psi^{10}$ has definite parity and
$\langle \Psi_R | {\bbox L}_T\cdot{\bbox \Omega}^-_T| \Psi_R\rangle=0$,
furnishing a simple example of a {\it racemic mixture}. (Notice the
similarity between our two-particle racemic mixture and a Dirac electron, which 
is made out of two Weyl spinors: 
$D^{(\frac{1}{2},0)}\oplus D^{(0,\frac{1}{2})}$ \cite{Ram89}.) 

To summarise: The work has discussed a theoretical approach to photo-absorption
spectroscopies in noncentrosymmetric systems. 

A symmetry analysis based on homogeneous Lorentz group has indicated that, 
in the presence of inversion-symmetry-breaking hybridisation, valence electrons 
develop an orbital anapole moment. The coupling 
of such a moment to the orbital angular momentum yields a set 
of microscopic effective operators, which identify the electronic properties
probed by x-ray and optical dichroism (E1-E2 and E1-M1 interferences). 

It has also been shown that orbital chirality (enantiomophism) 
is described by a two-electron $so(3,1)$ invariant (pseudoscalar), which is measured
by optical natural circular dichroism. 

Stimulating discussions with E. Katz are gratefully acknowledged.
\begin{table}
\caption{Polarisation responses of E1-E2 and E1-M1 interferences.}
\vspace{.1in}
\begin{tabular}{l}
\\
$iT^{(0)}_0 ({\bbox \epsilon}^*,{\bbox \epsilon}, \hat{\bf k})_1
=-iT^{(0)}_0 ({\bbox \epsilon},{\bbox \epsilon}^*, \hat{\bf k})_1
=-i/\sqrt{6}$,\\
\\
$iT^{(2)}_q ({\bbox \epsilon}^*,{\bbox \epsilon}, \hat{\bf k})_1
=-iT^{(2)}_q ({\bbox \epsilon},{\bbox \epsilon}^*, \hat{\bf k})_1
=-i/2[[{\bbox \epsilon}^*\otimes
{\bbox \epsilon}]^{(1)}\otimes\hat{\bf k}]^{(2)}_q$,\\
\\
$T^{(1)}_q ({\bbox \epsilon}^{\parallel}, \hat{\bf k})_2= 
T^{(1)}_q ({\bbox \epsilon}^{\perp}, \hat{\bf k})_2 =
-1/2\sqrt{\frac{3}{5}}\hat{\bf k}_q$,\\
\\
$T^{(2)}_q ({\bbox \epsilon}^*,{\bbox \epsilon}, \hat{\bf k})_2
=-T^{(2)}_q ({\bbox \epsilon},{\bbox \epsilon}^*, \hat{\bf k})_2
=\frac{\sqrt{3}}{2}[[{\bbox \epsilon}^*\otimes
{\bbox \epsilon}]^{(1)}\otimes\hat{\bf k}]^{(2)}_q$,\\
\\
$T^{(3)}_q ({\bbox \epsilon}^{\parallel}, \hat{\bf k})_2
=[[{\bbox \epsilon}^{\parallel}\otimes{\bbox \epsilon}^{\parallel}]^{(2)}
\otimes\hat{\bf k}]^{(3)}_q
\neq T^{(3)}_q ({\bbox \epsilon}^{\perp}, \hat{\bf k})_2$
\\
$=[[{\bbox \epsilon}^{\perp}\otimes{\bbox \epsilon}^{\perp}]^{(2)}
\otimes\hat{\bf k}]^{(3)}_q$
\\ \\
\end{tabular}
\label{TableI}
\end{table}
\end{document}